\documentclass{moriond}

\def\mco{\multicolumn}

\def\ra{\rightarrow}

\def\ko{K^0}

\def\be{\begin{equation}}
\def\ee{\end{equation}}
\def\bea{\begin{eqnarray}}
\def\eea{\end{eqnarray}}

\usepackage{amssymb,mathrsfs}
\usepackage{amsmath}
\usepackage{multirow}
\usepackage{float}
\newcommand{\Tr}[1]{\text{Tr}\left[#1\right]}
\newcommand{\abs}[1]{\left\lvert#1\right\rvert}

\newcommand{\etaee}{\eta_{ee}}

\newcommand{\Photo}{}

\begin{document}
\vspace*{4cm}
\title{Probing the heavy neutrino hypothesis}

\author{ Xabier Marcano }

\address{Departamento de F\'{\i}sica Te\'orica and Instituto de F\'{\i}sica Te\'orica UAM/CSIC,\\
Universidad Aut\'onoma de Madrid, Cantoblanco, 28049 Madrid, Spain}

\maketitle\abstracts{
There is a strong experimental program searching for massive sterile neutrinos. 
Here we focus on the heavy regime above the GeV scale, where their existence can be probed in either high-energy colliders or in high-precision facilities. 
We first review the current experimental status at colliders, showing that the LHC already improves LEP results for mixings to electrons and muons, while mixings to taus remain challenging at a hadron collider. 
We also discuss the importance of exploring both lepton number violating and conserving signals, as well as different flavor channels. 
Finally, we present the latest global analysis of electroweak precision and flavor observables, showing that the intensity frontier provides the strongest constraints for heavy neutrino masses above the electroweak scale. 
}

\section{Introduction}

Right-handed neutrinos, also known as sterile neutrinos or heavy neutral leptons (HNLs), are among the simplest yet best motivated new particles to extend the particle content of the Standard Model (SM). 
They are introduced by a plethora of models explaining neutrino mass generation, often in connection with other open problems in the SM. Unfortunately, their mass scale is usually unknown, motivating a strong experimental program covering all possible ranges. 

Here, we focus on the heavy neutrino hypothesis, when they are above the few GeV scale. This energy scale represents the threshold above which the new neutrinos cannot be produced in light meson or $\tau$ decays, so we enter the high-energy colliders territory. 
Colliders such LEP and LHC can efficiently produce heavy neutrinos below the electroweak (EW) scale, but the rates fall down for heavier masses, until they become too heavy to be produced. 
Nevertheless, their existence induces deviations from unitarity of the leptonic mixing matrix, so they can still be probed for at the intensity frontier via EW precision observables, universality ratios or charged lepton flavor violating (cLFV) processes.
In fact, and as we summarize here, direct searches at colliders provide the strongest bounds for heavy neutrinos from few GeV to the EW scale\,\cite{Abada:2022wvh}, above which indirect bounds from precision physics\,\cite{Blennow:2023mqx} dominate the current heavy neutrino landscape.

\section{Current LHC bounds for heavy neutrinos}

\begin{figure}[t!]
\begin{center}
\includegraphics[width=.5\textwidth]{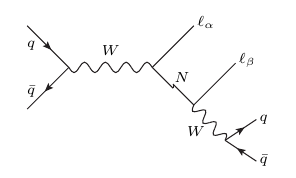}
\caption[]{Drell-Yann heavy neutrino production at hadron colliders leading to a dilepton signature. 
The thunder-shaped arrow indicates that the N could be of Dirac or of Majorana nature, short- or long-lived. }\label{DiagHNLatLHC}
\end{center}
\end{figure}

\begin{figure}
\begin{minipage}{\linewidth}
\centerline{\includegraphics[width=.9\linewidth]{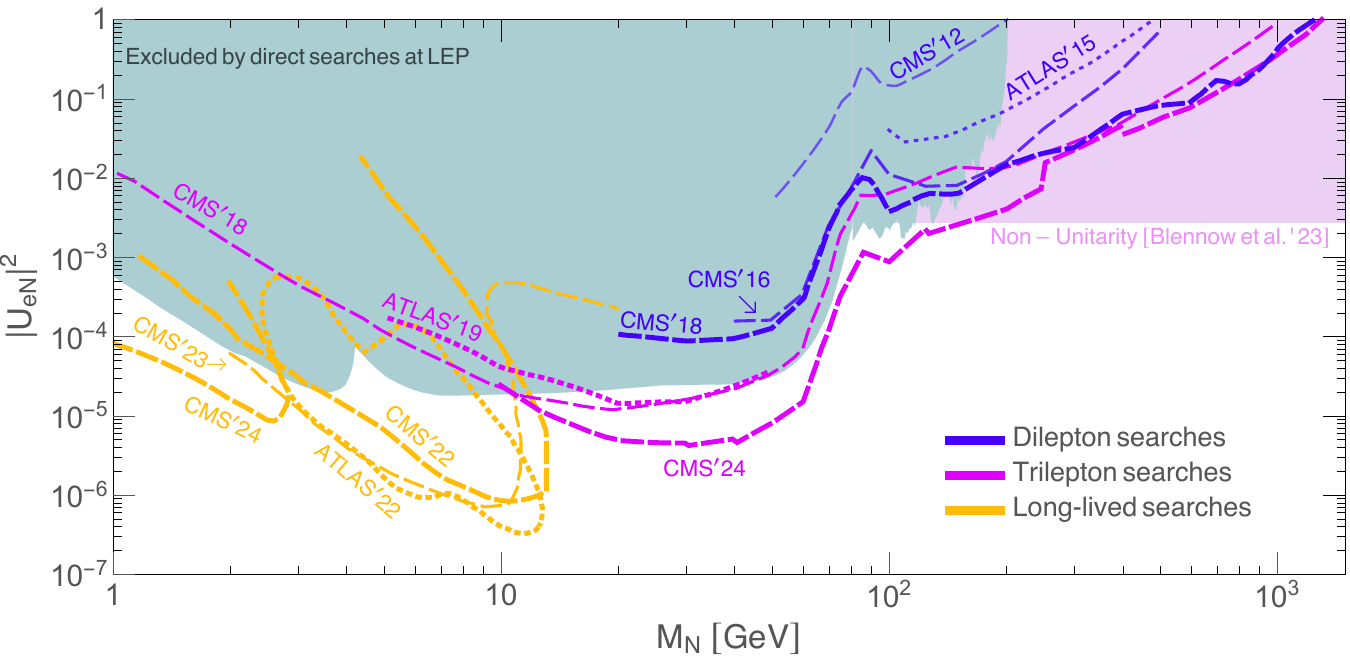}}
\centerline{\includegraphics[width=.9\linewidth]{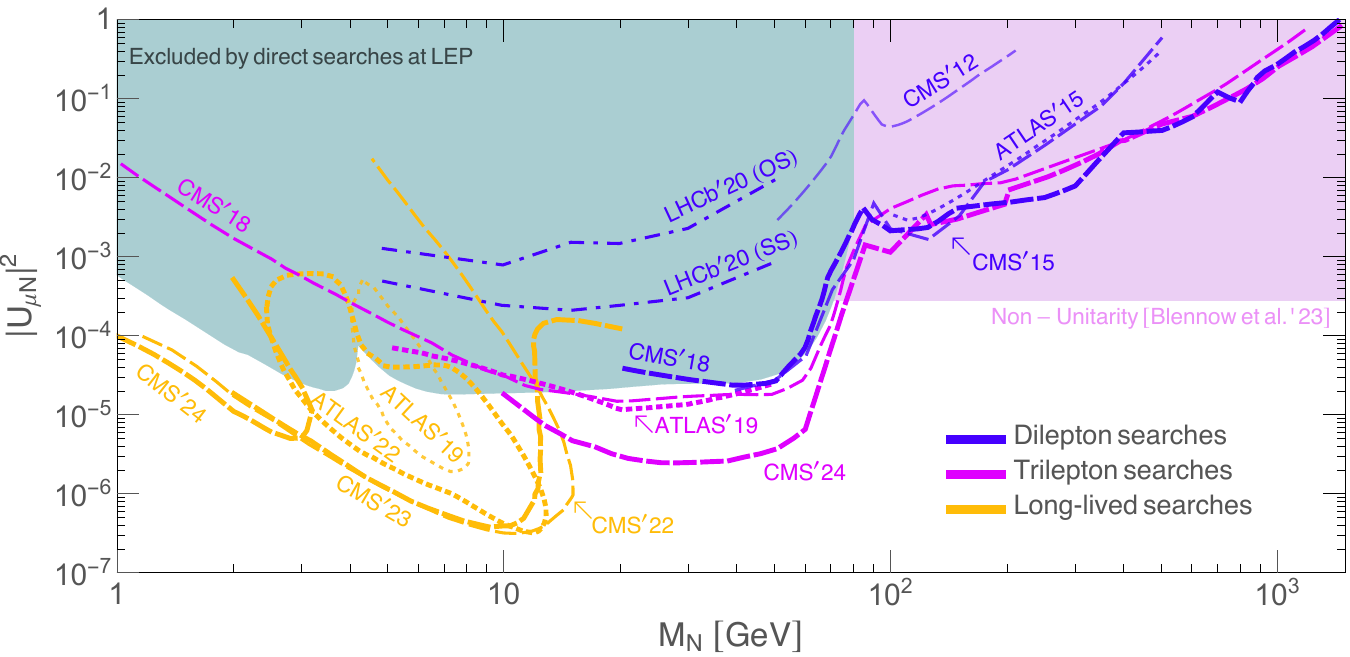}}
\centerline{\includegraphics[width=.9\linewidth]{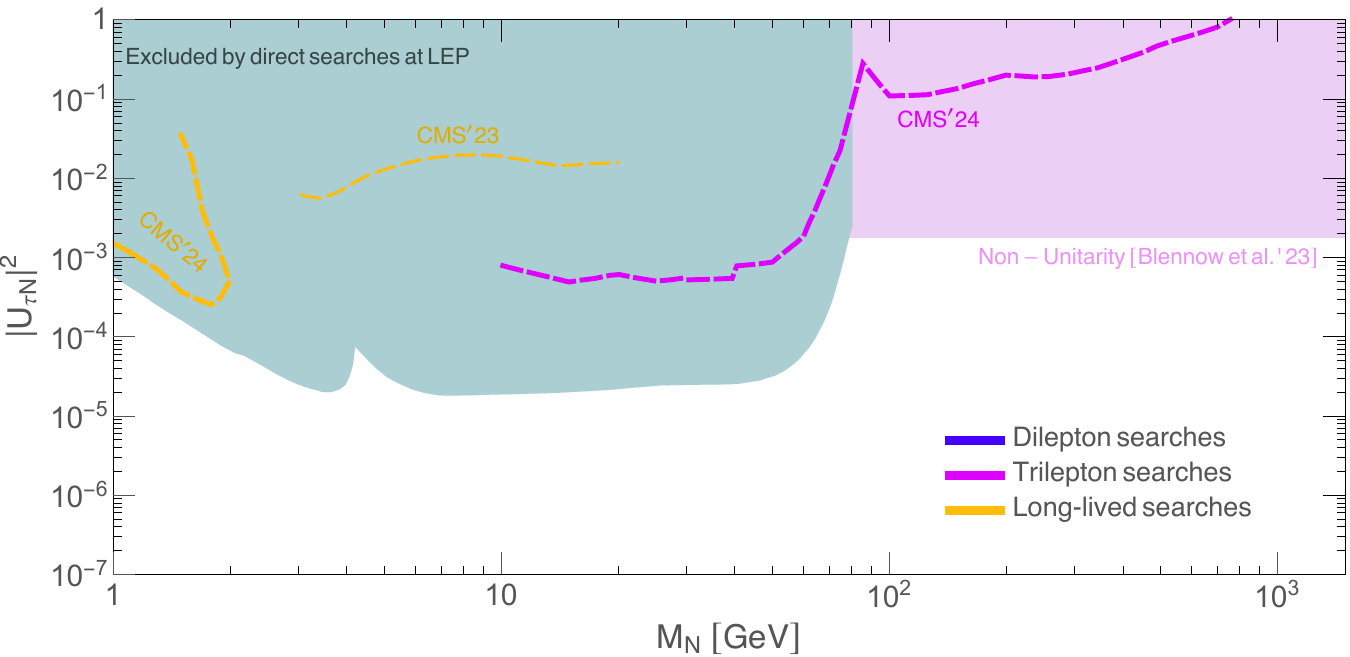}}
\end{minipage}
\caption[]{Summary of LHC searches for Majorana heavy neutrinos. Updated from {\it Abada et al.}\cite{Abada:2022wvh} with recent LHC results\,\cite{}. Included also the latest bounds from non-unitarity\,\cite{Blennow:2023mqx}, which dominate above the EW scale.}
\label{fig:finalbounds}
\end{figure}

We focus first on heavy neutrino masses $M_N$ ranging from few to hundreds of GeVs, which could be directly produced at high-energy colliders and their lifetimes are usually short enough to decay within the detectors, so we could discover them looking for their decay products. 

At a hadronic collider such as the LHC, the main production channel comes from Drell-Yan W or Z bosons and, for heavy masses, also from vector boson fusion channels such as $W\gamma$.
Notice however that current searches usually consider only the charged channel since the associated charged lepton helps triggering the event. 
Then, the $N$ will decay to charged leptons, neutrinos and quarks, leading to different signal topologies that we can classify according to the number of charged leptons as dilepton (as in Fig.~\ref{DiagHNLatLHC}) or tripletons. 
Moreover, depending on their masses and couplings, the heavy neutrino could be short or long-lived, so we can further split the signatures in prompt or displaced vertex signals.

We collect and classify in Figure~\ref{fig:finalbounds} all the heavy neutrino searches done at the LHC so far, shown in the usual mass-mixing plane.
This is an updated version of the figure by {\it Abada et al.}\cite{Abada:2022wvh}, where we have included the latest results from CMS\,\cite{CMS:2023jqi,CMS:2024ake,CMS:2024xdq}.
We did not include other recent results, such as the CMS long-lived search for very light masses\,\cite{CMS:2024ita} or the vector boson fusion searches\,\cite{CMS:2022hvh,ATLAS:2024rzi}.
The former gives the most relevant bounds for $M_N\lesssim2$~GeV, where there are stronger bounds from tau and light meson decays.
The latter are t-channel processes, so the heavy neutrinos are not {\it directly} produced as in the other searches, which consider on-shell neutrinos. Moreover, while this allows for sensitivities to very heavy masses, current bounds are several orders of magnitude weaker than those from precision data, as we will see later.

Figure~\ref{fig:finalbounds} reveals several interesting aspects. On the one hand, it makes manifest that the LHC is immersed in an intense program for heavy neutrino searches. On the other hand, it shows the complementarity between the different signal topologies to cover all the parameter space, especially between prompt and long-lived regimes. 
As a result, the LHC is already able to improve LEP results for all heavy neutrino masses and for mixings to electrons or muons. 
This is a non-trivial statement, since lepton colliders are extremely relevant for heavy neutrino searches and not easy to improve by a hadron collider beyond the fact that they can probe heavier masses. 
The reason is that LEP, and similarly future leptonic colliders such as the FCCee, combined a clean environment with a huge amount of Z bosons collection to search for HNLs produced in $Z\to\nu N$.
This kind of search, which is challenging to perform at the LHC~\footnote{Although proposals as the {\it inclusive displaced vertex searches}~\cite{Abada:2018sfh} could help.}, has the advantage of being equally sensitive to all neutrino mixings, explaining why the LHC is not able to improve LEP results for $U_{\tau N}$.

Another important remark about Figure~\ref{fig:finalbounds} is that it only shows the results assuming that the heavy neutrino is a Majorana particle. This is in fact the most frequent assumption, as it leads to lepton number violating (LNV) signatures that help to keep backgrounds under control. 
Unfortunately, current collider searches are sensitive only to relatively large mixings, too large to explain the masses of the light neutrinos unless a symmetry protected scenario is invoked. More specifically, this symmetry is an approximated conservation of lepton number, which also leads to the naive suppression of the expected LNV signal.
From this point of view, searching Dirac neutrinos via lepton number conserving (LNC) signals, such as the opposite sign dilepton search by LHCb\,\cite{LHCb:2020wxx}, or some of the latests analyses for long-lived or triplepton channels, seems more relevant to explore theoretically motivated scenarios. 
Nevertheless, a closer look to the cancelation of LNV signals in symmetry protected scenarios reveals that it is not as straightforward, and that it depends on the details and parameter space of the model.
In fact, studying both LNV and LNC signals at the same time could provide us information not only about the nature of the heavy neutrinos, but also about the neutrino mass generation, being able even to disentangle between different low-scale seesaw realizations\,\cite{Fernandez-Martinez:2022gsu}.
Therefore, we stress the importance of keep exploring both Majorana and Dirac hypotheses at colliders.

\section{Beyond minimal assumptions in LHC searches}

\begin{figure}
\begin{minipage}{\linewidth}
\centerline{\includegraphics[width=.49\linewidth]{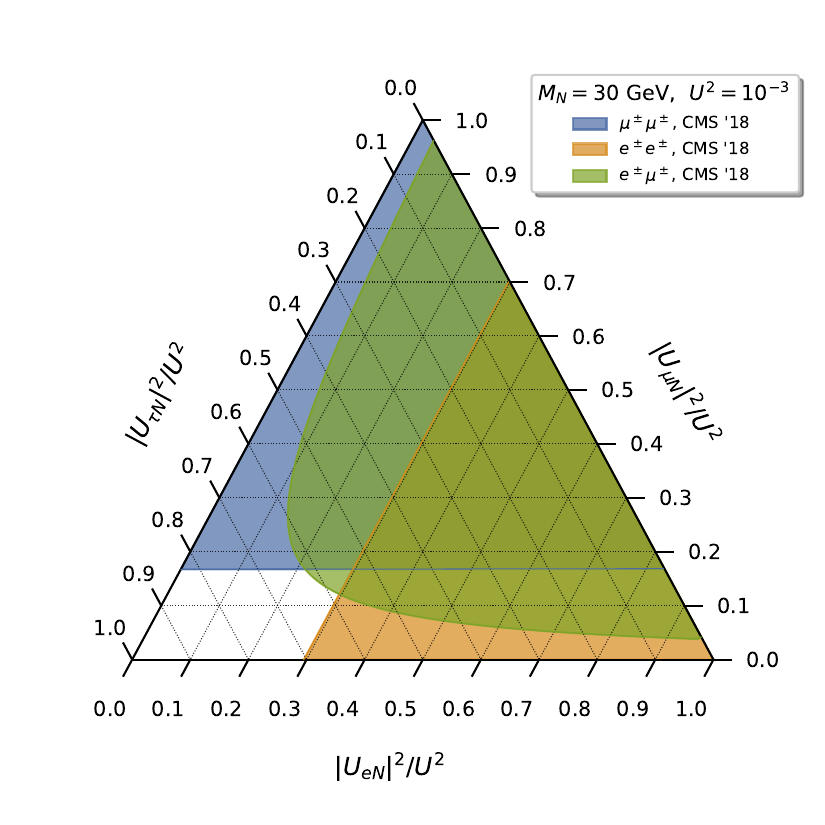}
\includegraphics[width=.49\linewidth]{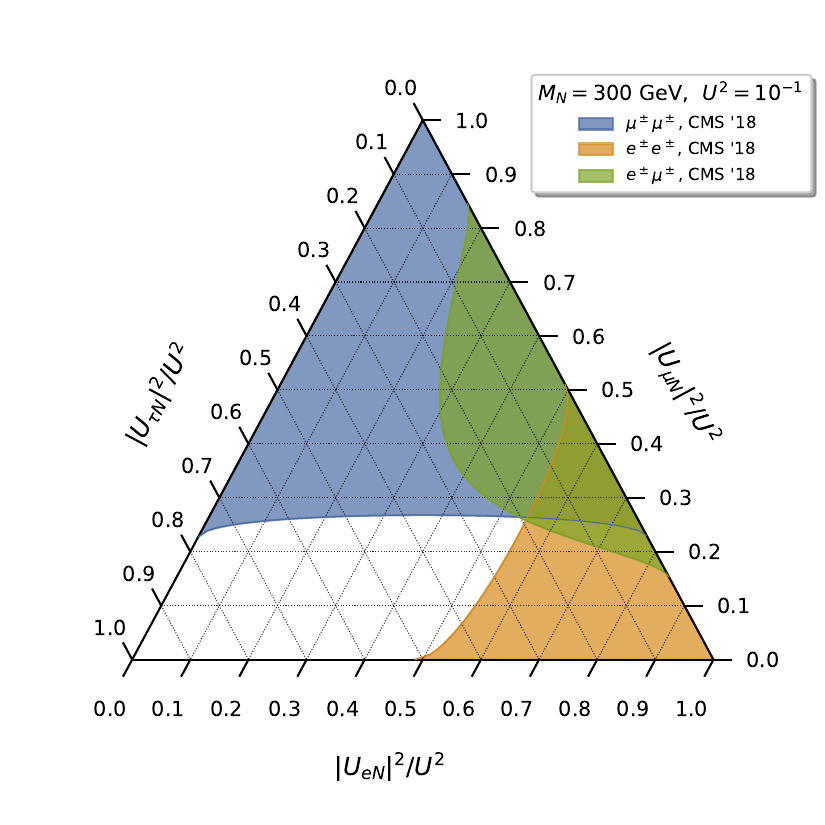}}
\end{minipage}
\caption[]{
Recast of same-sign dilepton searches by CMS to a general mixing pattern for two benchmark points of mass and total mixing $U^2\equiv|U_{eN}|^2+|U_{\mu N}|^2+|U_{\tau N}|^2$.
The white area is still allowed by LHC searches.}
\label{fig:ternaryFilled}
\end{figure}

With very few exceptions, the large amount of available heavy neutrino bounds have been derived relying on the assumption of the existence of a single (usually Majorana) neutrino that mixes with only one lepton flavor. However, most of the BSM scenarios involving new neutral leptons address the lepton mixing as a whole. The mixing pattern in these scenarios is expected to be quite complex, so applying the bounds derived in the context of simplified hypotheses does not seem adequate. Indeed, using these limits directly will in general overconstrain the parameter space\,\cite{Abada:2022wvh}. Consequently, most of the experimental bounds for heavy neutrinos need to be recast before being applied to a generic BSM scenario.

The main reason why we expect the bounds to be modified is due to the importance of the HNL decay width, which depends on every mixing $U_{\alpha N}$, and which plays a major role in the resonant searches we are interested in. This means that the final cross sections will depend on all of the mixings, even on those flavours that are not explicitly present in the charged leptons involved in the process. 
The generic flavor dependences of dilepton and trilepton channels at the LHC, assuming still a single heavy neutrino but with generic mixing patterns, was given by {\it Abada et al.}\cite{Abada:2022wvh}.
The main conclusion of that inspection of the different processes is that LHC results could be easily recast to generic mixing patterns if the bounds are provided for each of the flavor channels individually, besides the usually provided combination of channels that is performed to increase the final sensitivity. 

As an example, let us consider the same-sign prompt dilepton searches at CMS\,\cite{CMS:2018jxx}. This analysis provided individual bounds for the $ee$, $\mu\mu$ and $e\mu$ channels, considering each time the corresponding simplified mixing pattern. Nevertheless, given that we know how each of this individual channels depend on the different mixings\,\cite{Abada:2022wvh}, we can easily combine these results to cover all possible mixing patterns, as in the ternary in Figure~\ref{fig:ternaryFilled}.
On the other hand, trilepton searches, which currently give the best limits for most of the masses, are derived combining different flavor channels, since they all depend share the same $|U_{\alpha N}|^2$ dependency in the simplified scenarios ({\it e.g.} both $\mu\mu\bar\mu$ and $\mu\mu\bar e$ channels scale as $|U_{\mu N}|^2$ for a single Majorana neutrino that mixes only to muons). 
Nevertheless, as soon as we turn to a more complex scenario, each of these flavor channels will scale differently, and thus we cannot recast the bounds of the simplified scenario. 
Consequently, in order to constraint any generic BSM scenario, it is important to have access to the constraints imposed by each of the flavor channels individually (or even better to the correlations), not just the final combination.

\section{Current precision bounds for heavy neutrinos}

\begin{table}[t!]
\centering
    \begin{tabular}{|c|c|c|c|c|}
    \hline 
    &\multicolumn{2}{c|}{}&\multicolumn{2}{c|}{}\\[-2.5ex]
    \multirow{2}{*}{\bf G-SS}&\multicolumn{2}{c|}{LFC Bound} 
    &\multicolumn{2}{c|}{LFV Bound}\\[0ex]  
    \cline{2-5} &&&&\\[-2.5ex]
    & $68\%$CL& $95\%$CL
    & $68\%$CL& $95\%$CL \\
    \hline
    &&&&\\[-2ex]
    $\eta_{ee}$ &
    $[0.33, 1.0] \cdot10^{-3}$ & $[0.081, 1.4] \cdot10^{-3}$ & -&- \\[1.2ex]
    $\eta_{\mu\mu}$ & $1.5 \cdot10^{-5}$ & $1.4 \cdot10^{-4}$ & - &-\\[1.2ex]
    $\eta_{\tau\tau}$  & $1.6 \cdot10^{-4}$ & $8.9 \cdot10^{-4}$ &-&-\\[1.2ex]
    $\Tr{\eta}$ & $[0.28,1.2]\cdot10^{-3}$ & $2.1\cdot10^{-3}$ &-&-\\[1.2ex]
    $\abs{\eta_{e\mu}}$ & $1.4\cdot10^{-4}$ & $3.4\cdot10^{-4}$ & $\mathbf{8.4\cdot10^{-6}}$ & $\mathbf{1.2\cdot10^{-5}}$ \\[1.2ex]
    $\abs{\eta_{e\tau}}$ & $\mathbf{4.2\cdot10^{-4}}$ & $\mathbf{8.8\cdot10^{-4}}$ & $5.7\cdot10^{-3}$ & $8.1\cdot10^{-3}$ \\[1.2ex]
    $\abs{\eta_{\mu\tau}}$ & $\mathbf{9.4\cdot10^{-6}}$ & $\mathbf{1.8\cdot10^{-4}}$ & $6.6\cdot10^{-3}$ & $9.4\cdot10^{-3}$ \\[1.2ex]
    
    \hline    
    \end{tabular}
    \caption[]{Upper bounds (or preferred intervals) from non-unitarity effects in a general seesaw scenario.
    The LFC bounds are obtained from the global fit analysis to EW flavor and precision observables.
    For off-diagonal $\eta_{\alpha\beta}$ elements, we also derive limits from cLFV transitions and highlight the strongest bound for each flavor sector.
    See {\it Blennow et al.}\cite{Blennow:2023mqx} for details and other scenarios.
    }
    \label{table:GSSbounds}
\end{table}

Heavy neutrinos can also be probed at the intensity frontier since, even if they cannot be directly produced, their existence induces deviations from unitarity of the leptonic mixing matrix.
These deviations can be parametrized in general through a small Hermitian matrix $\eta$:
\begin{equation}\label{eq:eta}
N = (\mathbb{I}-\eta)\, U\,,
\end{equation}
where $U$ is the unitary matrix that diagonalizes the Weinberg operator. 
This parametrization is convenient as it encodes non-unitarity effects regardless of the UV completion that originates the deviations and how many fields it contains.
In the particular case of heavy neutrinos,  $\eta$ corresponds to (half of) the coefficient of the dim-6 operator obtained upon integrating them out, and therefore it is directly related to their mixings. 
Thus, setting bounds on $\eta$ implies setting bounds on the sum of the mixings of all the heavy neutrinos in the model.

Given the fact that the non-unitarity effects affect many low-energy processes, we have performed a global analysis to the most important EW precision and flavor observables. 
Our new analysis improves previous ones with a better handle of the correlations and the statistical treatment, as well as updating the list of observables. 
In particular, our fit includes: four determinations of the $W$ boson mass; two determinations of the effective weak mixing angle; five LEP observables at the $Z$-pole; a CMS measurement of the $Z$ invisible width; five weak decay ratios of lepton flavor universality; ten weak decays constraining the CKM unitarity; and cLFV observables. 
The results for the case of a general seesaw model (G-SS) are given in Table~\ref{table:GSSbounds}. 
All the details of the analysis, as well as the results for other well-motivated scenarios, can be found in {\it Blennow et al.}\cite{Blennow:2023mqx}.

These upper bounds can be understood as bounds on the total squared mixing (the sum) of all heavy neutrinos to a given flavor.
In the particular case considered before for LHC searches, with a single heavy neutrino or a single pair of pseudo-Dirac neutrinos, we have
\begin{equation}
\eta_{\alpha\alpha} = \frac12 \big|U_{\alpha N}\big|^2\,,
\end{equation}
so we can directly compare the limits from colliders and from precision observables, as done in Figure~\ref{fig:finalbounds} (and neglecting the mild preference for a non-zero $\etaee$). 
We see that non-unitarity effects provide the strongest bounds for heavy neutrinos heavier than the EW scale.

\section{Conclusions}

Heavy neutrinos can be searched for at both energy- and intensity-frontiers. 
In the former, the strong LHC program searching for heavy neutrinos is already providing the strongest bounds for electron and muon sectors, improving previous LEP limits thanks to the complementarity of the different kind of searches (mainly prompt and long-lived).
Exploring the tau sector at the LHC is more challenging, although the latest searches already provide the first LHC limits. 

We have also discussed the potential of collider searches in going beyond the usually considered simplified scenarios, stressing the importance of exploring both LNV and LNC signals to learn about the model behind the neutrino mass generation, and highlighting also the role of the different flavor channels in exploring the whole parameter space. 
In particular, we emphasized the fact that providing the exclusion limits from each of the flavor channels individually, or even better with correlations, would facilitate recasting the experimental results to generic and more realistic scenarios. 

Finally, we presented an updated and improved global analysis of EW flavor and precision observables, deriving the strongest bounds for heavy neutrinos with masses above the EW scale, even beyond the direct ones obtained by LHC. 
Therefore, we can summarize the current heavy neutrino landscape as being dominated by the high-energy frontier from the few GeV to the EW scale, above which we enter the intensity frontier territory.

\section*{Acknowledgments}
This project has received support from the European Union’s Horizon Europe Programme under the Marie Skłodowska-Curie grant agreement no.~101066105-PheNUmenal, and from the Spanish Research Agency (Agencia Estatal de Investigaci\'on) through the Grant IFT Centro de Excelencia Severo Ochoa No CEX2020-001007-S.

\end{document}